\documentclass[3p, twocolumn, 12pt, authoryear]{elsarticle} %
\usepackage{fancyhdr}

\journal{Atmospheric and Solar-Terrestrial Physics}

\makeatletter
\def\ps@pprintTitle{%
 \let\@oddhead\@empty
 \let\@evenhead\@empty
 \def\@oddfoot{\centerline{\thepage}}%
 \let\@evenfoot\@oddfoot}
\makeatother

\usepackage{mathtools}

\usepackage{amssymb}

\usepackage{hyperref}

\begin{document}

\thispagestyle{fancy}

\begin{frontmatter}



\title{On the alleged coherence between the global temperature and the sun's movement}



\author{Sverre Holm}
\address{Department of Informatics, University of Oslo, Norway}


\begin{abstract}

It has recently been claimed that there is significant coherence between the spectral peaks of the global temperature series over the last 160 years and those of the speed of the solar center of mass at periods of 10-10.5, 20-21, 30 and 60-62 years. Here it is shown that these claims are based on a comparison between spectral peaks in spectral estimates that assume that the global temperature data contains time-invariant spectral lines. However, time--frequency analysis using both windowed periodograms and the maximum entropy method shows that this is not the case. An estimate of the magnitude squared coherence shows instead that under certain conditions only coherence at a period of  15-17 years can be found in the data. As this result builds on a low number of independent averages and also is unwarranted from any physical model it is doubtful whether it is significant.
\end{abstract}


\end{frontmatter}


\section{Introduction}
\label{}


Scafetta has argued that there is a strong coherence between the global temperature series over the last 160 years and the movement data of the sun around the barycenter (center of mass) of the solar system. This claim is based on coincidence between spectral peaks of the two time series. Such a coherence would, if proven to be correct, go  a long way towards showing that there is a celestial origin of the climate oscillations on the 10-62 year scale. 

The claimed coherence is presented in \citet{Scafetta2010} and it depends strongly on the result of maximum entropy spectral analysis of the climate series and the identification of specific peaks in the spectra. In that paper a total of 11 peaks with periods ranging from 6 years and above are identified and approximately matched with a planetary or moon--earth oscillation. 

In a follow-up work, only the five main periodicities from period 9 years and above are discussed \citep{Scafetta2012shared}. We will not discuss the 9-year periodicity claimed to correlate with the lunar orbit here, so the four periodicities which are claimed to have an origin in the solar orbit are at 10-10.5, 20-21, 30 and 60-62 years.

The underlying hypothesis of these papers is that there are periodicities in the climate series which correspond to periodicities in the solar system where the movement of the sun around the barycenter is used as a proxy. The logical consequence is drawn from this that the better the climate series periodicities can be understood and in particular the better their origin can be understood, the more they can be be used for predicting future climate development. This last point is elaborated in the papers \citep{Scafetta2012b,solheim2012long} which will not be discussed here.

The purpose of this paper is to reanalyze the data for coherence. The main tools are time--frequency spectral analysis and the magnitude squared coherence. 

\section{Analysis Methods}

\subsection{The maximum entropy method}

Maximum entropy spectral analysis  is a high resolution method which fits an all-pole spectral model to the data. It is able to bring out spectral peaks that many other methods cannot show. One of the the most used numerical implementations is due to \citet{burg1967maximum}. During the 70's the Burg method was investigated for its properties and many applications were explored. One example application which we have been involved in is ocean wave spectral analysis \citep{Holm1979}. 

When the maximum entropy method (MEM) is applied one needs to be aware of its properties and in particular some pitfalls. One of them is that if a too high order is used, the spectral peaks will become sharper and sharper and eventually spurious peaks will appear. \citet{ghil2002advanced}, the source for the analysis methods of \citet{Scafetta2010}, also warn against using a too high order saying that an upper bound is generally taken as half the number of samples. An even stricter criterion is the recommendation that the maximum order should be between one-third and one-half of the number of samples \citep{kay1979effects}.

A second property is that too high order increases the chance that spectral peaks may split into two \citep{fougere1976spontaneous}. Third, there can be bias in the frequency location of peaks especially for frequencies where only a few periods are contained in the time series. The bias may depend on initial phase \citep{chen1974experiments}.

Fourth, \citet{ghil2002advanced} caution against applying MEM uncritically to nonstationary time series or data which are not well approximated by an AR process and advice that cross testing with other spectral analysis methods then is particularly important. This is particularly relevant here as the periodicities will turn out to be time-varying in the climate series data. 

Although MEM is a high resolution spectral analysis method, high resolution is not the same as high accuracy. Resolution shows how well two lines can be distinguished from each other. This is not so much of an issue for the data analyzed here. Therefore we will rely to a large extent on the periodogram and following \citet{ghil2002advanced} will validate MEM against the windowed periodogram, although MEM will have finer widths and seemingly better accuracy. 

\subsection{The windowed periodogram}

The periodogram has the desirable property that for a single sinusoid or for well isolated sinusoids in noise, it is the maximum likelihood estimator of frequency \citep{kay1988modern}. But one needs to be cautious in applying the periodogram as both the length of the window and the choice of a window function may be crucial for good performance. This is especially the case here where there may be very strong spectral peaks that can leak into and obscure other interesting parts of the spectrum. Therefore a Kaiser window is used here since its performance can be adjusted by a single parameter, $\beta$. Note that Kaiser's original definition uses the $\alpha=\beta/\pi$ parameter \citep{kaiser1974nonrecursive}, while we follow the convention in Matlab (\textregistered Mathworks Inc., Natick, MA) which characterizes the window by $\beta$. The larger the value, the lower the sidelobes and the poorer the frequency resolution will be. Prior to windowing, the data is detrended by removing any linear trend. The main purpose of that is to remove any zero-frequency components that may leak into the interesting high period part of the spectrum.

Another consideration is that nonlinear processing of the amplitude of the spectral estimator may make features more visible. \citet{kay1984high} show how one spectral estimator with what seems to be low resolution may be transformed into another which in the display seems to have higher resolution. The transform is a monotonic transform. Inspired by this, we have used a power law function to transform the periodogram into a visually as attractive plot as possible using the transformation:
\begin{equation}
T(P_{xx}(f)) = P_{xx}(f)^{a},
\label{eq:fft}
\end{equation}
where $P_{xx}$ is the power spectrum estimate and $0<a\le1$. Only the periodogram output is transformed in this way, the Burg MEM result is unprocessed ($a=1$) and converted to dB.

\subsection{Time-Frequency Analysis}

It will be clear later that the global temperature data in particular is not time-invariant. As spectral estimation methods that assume stationarity will produce many spurious peaks when applied to nonstationary data, this is the main criticism of the results of \citet{Scafetta2010}.

In the time--frequency analysis, we have used Burg's Maximum Entropy Method and the windowed periodogram with an offset of 1 year between each analysis window. The length of the data window must be chosen and the longer the window, the better the frequency resolution. On the other hand, the shorter the window, the better the ability to track time variations in the data. For the examples shown here, a window length of 60 years was found to be a reasonable compromise. 

Such a compromise between time and frequency resolution is the background for the development of multi-resolution methods such as wavelets. They often have poorer frequency resolution and better time resolution at high frequencies compared to the low frequencies. We have chosen not to use wavelets here, primarily because it is not used in \citet{Scafetta2010,Scafetta2012shared}, but also because it may often be hard to understand and evaluate the performance and accuracy of these methods, such as the time- and frequency-resolution in the various frequency bands. This may also make it hard to separate features of the data from properties of the analysis method.

The order used for MEM analysis is found by trial and error and comparison with the periodogram as advised in  \citet{ghil2002advanced}, keeping in mind that it should not exceed between $1/3$ and $1/2$ of the number of samples and that the window length should not smooth out nonstationarities. Based on a number of analyses where Figs.\ 3 and 6A of \citet{Scafetta2010} were reproduced, we ended up choosing order 300 for a time series of length 60 years. This is $300/(12\cdot 60) \approx 42\%$ of the data length which is somewhat more conservative than the order 1000 in \citet{Scafetta2010} which is equivalent to $1000/(12 \cdot 160) \approx 52\%$. Such a high order in combination with a window that spans the entire data record of 160 years contributes to additional peaks in the spectra shown in Figs.\ 3 and 6A of \citet{Scafetta2010}. Although in a later paper Scafetta has realized that half the length of the record is the highest allowed pole order for MEM,  \citet{scafetta2012multi}, there is still no discussion of the smearing effect of analyzing the whole record in one analysis window.

\subsection{Magnitude squared coherence estimation}
The magnitude squared  coherence (MSC) is a well-established correlation factor in the frequency domain. It has been used for decades as the best method for identifying coupling between signals \citep{bendat1980}. The MSC is defined as the normalized squared cross spectrum between two time series $x(t)$ and $y(t)$

\begin{equation}
C_{xy}(f) = \frac{|P_{xy}(f)|^2}{P_{xx}(f)\cdot P_{yy}(f)},
\label{Eq:msc}
\end{equation}
where $P_{xx}(f)$ and $P_{yy}(f)$ are the power spectra of the two time series and $P_{xy}(f)$ is the complex cross spectrum. $C_{xy}(f)$ will always be between 0 and 1. In order to estimate it, both time series have to be divided into several weighted and partly overlapping segments which are Fourier transformed and averaged.

The estimator finds relationships between data based on linear systems theory where the effect, $y(t)$, is a filtered version of the cause, $x(t)$, with noise added. The smaller the contribution of the noise at a particular frequency, the closer the MSC will be to 1. If on the other hand there is a chaotic system coupling between input and output, the MSC will not be able to discriminate between that and additive noise. In such cases it is not so useful. However, there is nothing in \citet{Scafetta2010} that indicates chaotic coupling, so the MSC should be an appropriate measure to use.

Statistics of the estimator is given in \citet{carter1973statistics} which says that the maximum bias, the maximum standard deviation, and the maximum rms error only decay as $N_d^{-1/2}$, where $N_d$ is the number of disjoint segments or independent averages. If only one segment is used to cover the whole length of the time series, the estimator will always give $C_{xy}(f)=1$ independent of the data and thus the estimate is useless. The maximum bias is shown by \citet{carter1973statistics} to be approximately bounded by $0.5 \sqrt{\pi/N_d}$ and the largest bias occurs for values of MSC close to 0. The number of independent averages should therefore be large in order to get a good estimate, on the order of $N_d=10$ or more. 

Since a Kaiser window which tapers the ends heavily down is used, 75\% overlap is used between the $N$ segments. The number of segments with this overlap is:

\begin{equation}
N = 1 + \lfloor{\frac{N_{tot}-N_{seg}}{0.25 N_{seg}}}\rfloor
\label{Eq:N}
\end{equation}
where $N_{tot}$ is the total number of monthly samples and $N_{seg}$ is the analysis interval, e.\ g.\ $N_{seg}=40\cdot12$ for a 40 year interval.

The  actual value for the number of independent estimates, $N_d$, will be less than $N$ and is estimated from the correlation between the data windows at the 50\% and 75\% points ($c(0.5)$ and $c(0.75)$) as given by \citet{harris1978use}:

\begin{equation}
N_d \approx N/[1+2 c^2(0.75) + 2c^2(0.5)]
\label{Eq:Nd}
\end{equation}


Confidence intervals for the MSC have been estimated using the method of \citet{wang2004exact}. The independence threshold has also  been estimated using the method of \citet{carter1987coherence} and \citet{wang2004optimising}. Coherence estimates below this value indicates that the two processes are independent. It is:

\begin{equation}
IT = 1-(1-P_d)^{1/(N_d-1)}
\label{Eq:Indep}
\end{equation}
where $P_d$ is the confidence level, $0.95$ or $0.99$ in our case.

It should be mentioned that the coherence measure used here is different from the measure mentioned in  \citet{Scafetta2010,Scafetta2012shared}. In the first paper, although a term which is called coherence is listed and discussed, there is little documentation for how it is estimated beyond a comparison of spectral peaks. The coherence measure in \citet{Scafetta2012shared} is based on the joint power statistics or the cross-spectrum. Therefore it is close to the numerator  of Eq.\ (\ref{Eq:msc}). However, without the normalization by the power spectra in the denominator, this measure may indicate large peaks where there is large energy, regardless of whether there is spectral correlation or not. Thus it is not a good indicator of coherence.

\subsection{Implementation}
Only standard functions of Matlab\textregistered have been used. The MEM-method is \texttt{pburg.m}, the periodogram is estimated by using \texttt{periodogram.m}, the window function is \texttt{kaiser.m}, and detrending is performed by using \texttt{detrend.m}. The function \texttt{mscohere.m} is used for magnitude squared coherence estimation.

The number of samples in the frequency domain was set to a very large number ($2^{16}= 65536$) in order to interpolate the spectral plots and avoid sampling artifacts. With monthly sampling, the maximum frequency supported is 6 year$^{-1}$, while the periods of interest between 60 and 6 years are in the frequency range from $60^{-1} \approx 0.017$ to $6^{-1} \approx 0.17$ year$^{-1}$. Thus only the $6^{-1}/6 \approx 2.8 \%$ lowest part of the spectrum is really of interest. Despite the interpolation there will be only about $65536 \cdot 2.8\%/2 \approx 918$ samples in this interval, which is a reasonable number. 


\section{Data Analysis}

\subsection{Temperature data}

The data is the monthly sampled global surface temperature (HadCRUT3) series from 1850 to 2010.75 ($N_{tot}=1929$ samples). The dataset is a combination of land and marine temperature anomalies on a $5^{\circ}$ by $5^{\circ}$ grid-box basis, see plot in Fig.\ 1 of \citet{Scafetta2010} and description in \citet{brohan2006uncertainty}. 

\begin{figure}[tb]
\begin{center}
	\includegraphics[width=.9\columnwidth]{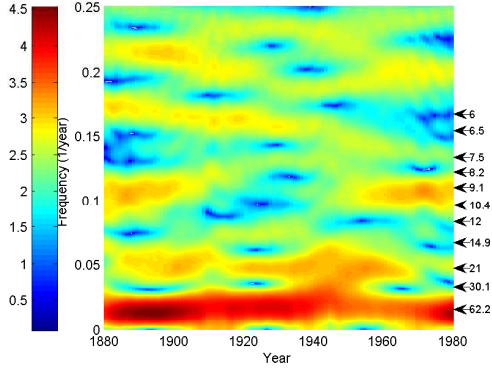} 
	\caption{Periodogram-based temperature spectrogram, 60 year window and Kaiser window with $\beta=3$ and a shift of 1 year between windows, $a=0.4$. Period in years as estimated by \citet{Scafetta2010} shown with arrows on right-hand side.}
	\label{fig:Temp-spectrogram60}
\end{center}
\end{figure}

\begin{figure}[tb]
\begin{center}
	\includegraphics[width=.9\columnwidth]{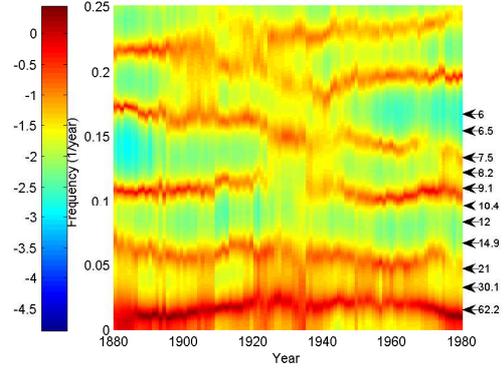}
	\caption{MEM-based temperature spectrogram, 60 year window, order 300, shift 1 year between windows. Period in years as estimated by \citet{Scafetta2010} shown with arrows on right-hand side.}
	\label{fig:Temp-spectrogramMEM60}
\end{center}
\end{figure}

The time--frequency spectrogram of the temperature data is shown in Figs.\ \ref{fig:Temp-spectrogram60} and \ref{fig:Temp-spectrogramMEM60}. The periodogram is transformed with $a = 0.4$ in Eq.\ (\ref{eq:fft}) and the MEM estimate uses order 300. Both use a window length of 60 years which is a length that does not seem to smear the nonstationarity of the data too much. It should be noted that there is more temporal variation in these data than in e.g.\ sunspot data, so a shorter data window than that used for MEM analysis of sunspot data \citep{currie1973fine} must be used. A Kaiser window with $\beta=3.0$ was used for the periodogram. This is a mild form of windowing with first sidelobe at $-23.8$ dB. Both figures show in general the same result, although the MEM plot has sharper and sometimes more continuous lines.

The 11 periodicities claimed by \citet{Scafetta2010} are denoted by arrows in the margin. From the time--frequency spectrogram only three of these can be recognized with some confidence. The strongest one is at about 60 years and is close to time-invariant. It is in agreement with the 65-70 year cycle reported by \citet{schlesinger1994oscillation}. There is also a line in the 15-20 year range which varies with time as well as one around 9 years which comes and  goes and varies in frequency.


\subsection{Solar movement data}

The solar orbit data has been taken from the Horizons system of JPL  by generating equal spaced monthly data between 15 January 1750 and 16 December 2100. Prior to this analysis the data set was confirmed to reproduce Fig.\ 4 and the spectrum of Fig.\ 6A of \citet{Scafetta2010}.

\begin{figure}[bt]
\begin{center}
	\includegraphics[width=.9\columnwidth]{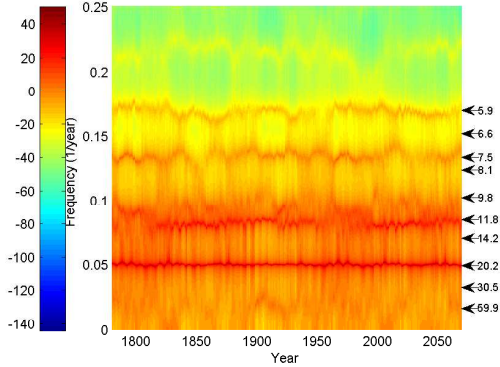}
	\caption{MEM-based spectrogram of the SCMSS (speed of the center of mass of the solar system), 60 year window, order 300, shift 1 year between windows. Period in years as estimated by \citet{Scafetta2010} shown with arrows on right-hand side.}
	\label{fig:SCMSS-spectrogramMEM60}
\end{center}
\end{figure}

The MEM spectrogram for speed of the center of the mass of the solar system (SCMSS) data is shown in Fig.\ \ref{fig:SCMSS-spectrogramMEM60} with a window length of 60 years. The 10 periodicities of \citet{Scafetta2010} are denoted by arrows in the margin. Due to the near time-invariance of this dataset, several of these are also visible in the time--frequency spectrogram. 

The strongest stationary component is that at about 20 years which can be explained by the period for which the position of the Sun, Jupiter, and Saturn are re-aligned (the synodic period of Jupiter and Saturn). There is also a fairly consistent periodicity at 11.9 years which is the revolution period of Jupiter. There are weaker almost constant-frequency lines at 9.8, 7.5, and 5.9 years also.


Although there is some overlap between the frequencies of the main lines of Figs. \ref{fig:Temp-spectrogramMEM60} and \ref{fig:SCMSS-spectrogramMEM60}, the time-varying nature of the lines in the temperature data is quite different from the almost constant frequencies in the SCMSS data. The importance of these differences will be even clearer when the magnitude squared coherence is estimated.

\subsection{Coherence Analysis}
The magnitude squared coherence (MSC) is estimated by averaging the Fourier transforms of windowed, overlapped segments of the data. As the SCMSS data has strong periodicities it is necessary to use a Kaiser window with $\beta=7$, i.e.\ with a first sidelobe as low as $-51$ dB in order to avoid spectral leakage. This can also be seen from  the large dynamic range of Fig.\ \ref{fig:SCMSS-spectrogramMEM60} compared to that of Fig.\ \ref{fig:Temp-spectrogramMEM60}.  Due to the strong tapering effect, 75\% overlap is used for the MSC analysis. 

\begin{figure}[tb]
\begin{center}
	\includegraphics[width=.9\columnwidth]{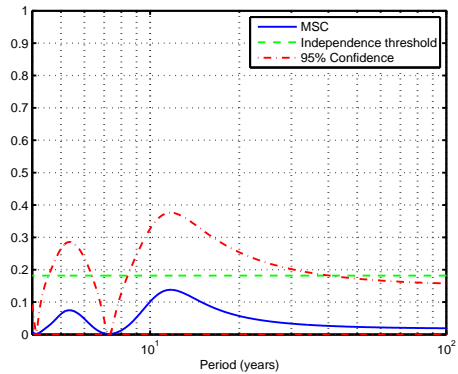}
	\caption{Magnitude squared coherence between speed of center of mass of the solar system (SCMSS) and global temperature data (solid line) with segment length 20 years. The 95\% confidence intervals (dash-dot lines) and the 95\% independence threshold of 0.18  (dashed line) are also shown. Peaks are at 5.3 and 11.7 years.}
	\label{fig:MSC-SCMSS-Temp}
\end{center}
\end{figure}

First an analysis is done over windows of length 20 years. Over an interval of 160 years this gives $N=29$ analysis windows, Eq.\ (\ref{Eq:N}). Using the properties of the window and the overlap, Eq.\ (\ref{Eq:Nd}) gives as a result that there are $N_d=15.9$ degrees of freedom in the estimate of MSC with this particular window. This gives an independence threshold of $IT=0.18$ at the 95\% confidence level and $IT=0.27$ at the 99\% level. The coherence between the solar movement (SCMSS) and the global temperature data in Fig.\ \ref{fig:MSC-SCMSS-Temp} is not above 0.15 for any frequency. This is a very low value which is below the independence thresholds. 


The analysis was then repeated with segment lengths of 30, 40, and 50 years. The resulting decrease in the number of independent estimates, $N_d$, the position and the peak value of the magnitude square coherence and the independence threshold at the 95  and 90\% confidence levels are shown in Table \ref{table:MSC}. 

\begin{table}[t]
\footnotesize
\centering
\begin{tabular}{c | c c c c c c} 
$N_{seg}/12$  & $N$ &$N_d$ & Peak    & Peak & IT & IT\\
(years)     &        &     & (years)&  MSC & 95\%        & 99\% \\    
\hline
20  & 29 & 15.9 & 11.7 & \bf{0.14} & 0.18 & 0.27  \\
30  & 18& 9.9   & 14.9 & \bf{0.35} & 0.29 & 0.40  \\
40  &13 &7.1   & 16.8 & \bf{0.59} & 0.39 & 0.53 \\
\hline
\end{tabular}
\caption{Number of averages $N$, number of independent averages $N_d$, MSC peak location and value, and independence thresholds (IT) at the 95\% and 99\% confidence levels  as a function of segment length.}
\label{table:MSC}
\end{table}



\begin{figure}[tb]
\begin{center}
	\includegraphics[width=.9\columnwidth]{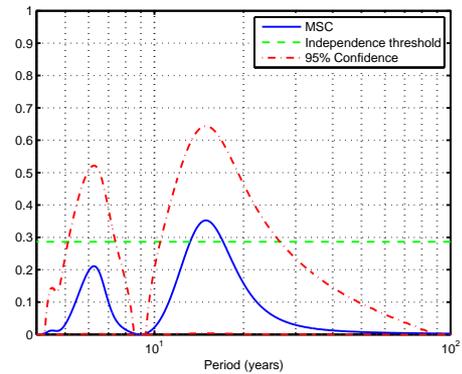}
	\caption{Magnitude squared coherence between speed of center of mass of the solar system (SCMSS) and global temperature data (solid line) with segment length 30 years. The 95\% confidence intervals (dash-dot lines) and the 95\% independence threshold of 0.29 (dashed line) are also shown. Peaks are at 6.3 and 14.9 years.}
	\label{fig:MSC-SCMSS-Temp-30}
\end{center}
\end{figure}

The maximum MSC-value increases with window length, but so does the uncertainty as the number of independent averages, $N_d$, decreases. When 30 year windows are used with the same data window and the same overlap there is only $N_d=9.9$ degrees of freedom and the independence threshold increases to $0.29$ at the 95\% confidence level and $0.4$ at the 99\% level, see Fig.\ \ref{fig:MSC-SCMSS-Temp-30}. In this case there is a peak in the coherence at $14.9$ years with a value of $0.35$ which is slightly above the 95\% confidence level, but below the 99\% confidence level.


The value increases when 40 year windows are used and there is a peak in the coherence at $16.8$ years which is above the independence thresholds, see Table \ref{table:MSC}. However, as the number of independent averages is only $N_d=7.1$ the reliability of this result is less than those for the lower window lengths.


\section{Discussion and Conclusion}
\citet{Scafetta2010} claimed the global temperature series for the last 160 years to have  spectral lines at 21, 30 and 62 years. Time--frequency analysis shows that the lines are time-varying (Figs.\ \ref{fig:Temp-spectrogram60} and \ref{fig:Temp-spectrogramMEM60}) and very different from the nearly constant lines in the time--frequency plot for the speed of the center of mass of the solar system (SCMSS) (Fig.\ \ref{fig:SCMSS-spectrogramMEM60}).  The supposed periodicity around 30 years in \citet{Scafetta2010} is not really present in the climate series at all and could be an artifact due to a combination of model overfitting and smearing due to the time-invariance assumption which has been forced on the data. The claimed spectral peaks by \citet{Scafetta2010} for the global temperature series are therefore not reproducible if proper consideration is taken of the time-varying nature of the data. 
The only significant coherence between the climate series and the sun's movement that was possible to find was at 15-17 years (Table \ref{table:MSC}). However, both the low number of independent averages that it builds on as well as the lack of a physical explanation for this coherence, makes us hesitate to claim that it is significant. 


\section{Acknowledgement}

I would like to thank Prof.\ ShouYan Wang, Suzhou Institute of Biomedical Engineering and Technology for stimulating discussions and for sharing his code for confidence interval estimation for the MSC. I also thank  Prof.\ F.\ Albregtsen and Dr.\ A.\ Austeng, University of Oslo, as well as Prof.\ K.\ Rypdal and Dr.\ M.\ Rypdal, University of Troms\o, Norway for helpful discussions during the work with this paper.

%




\bibliographystyle{elsarticle-harv}

\bibliography{../klima}







\end{document}